\documentclass[twocolumn,english,aps,prb,superscriptaddress,bibnotes,amsmath,amssymb,floatfix,nobalancelastpage]{revtex4-2}
\usepackage[colorlinks=true,citecolor=blue,linkcolor=magenta]{hyperref}
\usepackage[utf8]{inputenc}
\usepackage[english]{babel}
\usepackage{amsmath,amsfonts,amssymb}
\usepackage[T1]{fontenc}
\usepackage{url}
\usepackage{soul}
\usepackage{changes}
\usepackage{amsmath}
\usepackage{amsfonts}
\usepackage{amssymb}
\usepackage[version=3]{mhchem}
\usepackage{stmaryrd}
\usepackage{epstopdf}
\usepackage{graphicx}
\usepackage{lettrine}

\newcommand{\fref}[1]{Fig.~\ref{#1}}

\begin{document}

\title{Photonic Radar for Contactless Vital Sign Detection}
\author{Ziqian Zhang}
\affiliation{Institute of Photonics and Optical Science (IPOS), School of Physics, The University of Sydney, NSW 2006, Australia}
\affiliation{The University of Sydney Nano Institute (Sydney Nano), The University of Sydney, NSW 2006, Australia}
\author{Yang Liu}
\email{yang.liu@sydney.edu.au}
\affiliation{Institute of Photonics and Optical Science (IPOS), School of Physics, The University of Sydney, NSW 2006, Australia}
\affiliation{The University of Sydney Nano Institute (Sydney Nano), The University of Sydney, NSW 2006, Australia}
\author{Tegan Stephens}
\affiliation{Bird and Exotics Veterinarian, NSW 2017, Australia}
\author{Benjamin J. Eggleton}
\email{benjamin.eggleton@sydney.edu.au}
\affiliation{Institute of Photonics and Optical Science (IPOS), School of Physics, The University of Sydney, NSW 2006, Australia}
\affiliation{The University of Sydney Nano Institute (Sydney Nano), The University of Sydney, NSW 2006, Australia}
\date{\today}
\maketitle
\noindent \textbf{Vital sign detection is used across ubiquitous scenarios in medical and health settings. Contact and wearable sensors have been widely deployed. However, they are unsuitable for patients with burn wounds or infants with insufficient attaching areas. Contactless detection can be achieved using camera imaging, but it is susceptible to ambient light conditions and creates privacy concerns. Here, we report the first demonstration of a photonic radar for non-contact vital signal detection to overcome these challenges. This photonic radar can achieve millimeter range resolution based on synthesized radar signals with a bandwidth of up to 30 GHz. The high resolution of the radar system enables accurate respiratory detection from breathing simulators and a cane toad as a human proxy. Moreover, we demonstrated that the optical signals generated from the proposed system can enable vital sign detection based on light detection and ranging (LiDAR). This demonstration reveals the potential of a sensor-fusion architecture that can combine the complementary features of radar and LiDAR for improved sensing accuracy and system resilience. The work provides a novel technical basis for contactless, high-resolution, and high-privacy vital sign detection to meet the increasing demands in future medical and healthcare applications.
}

\noindent Vital signs -- a group of clinical measurements reflecting the essential body functions -- are used as diagnostic parameters for monitoring medical and health conditions. Vital sign detection is widely employed across ubiquitous scenarios, such as intensive care units (ICUs) for patients with critical health conditions, day-and-night health monitoring in aged care facilities to prevent unattended medical emergencies, and vehicles to determine the occurrence of drivers' drowsiness \cite{Cardillo2020, Dunn2021b}. Conventional vital sign detection relies on contact-based devices, such as pulse oximeters that use electrodes to detect weak electrical changes as a consequence of cardiac contractions (electrocardiography, ECG) and smartwatches based on the intensity variation of infrared probe light caused by blood flow and volume changes (photoplethysmography, PPG) \cite{Ren2017}. Although widely deployed, contact-based methods can cause discomfort for round-the-clock monitoring \cite{Mercuri2017, Massaroni2020, Presti2020}. Despite improved user experience for wearable sensors in bands or clothes, they are unsuitable for patients with burn wounds, skin irritations, or infants with insufficient attaching areas \cite{Liu2019}. Non-contacting methods based on optical sensors have been explored, for instance, using cameras to track certain body regions of interest \cite{Bennett2017, Chaichulee2017, Lyra2021}. However, camera-based systems create privacy concerns. Moreover, these optical approaches are sensitive to ambient light conditions and thermal interference.

\begin{figure*}[ht!]
 \centering{
 \includegraphics{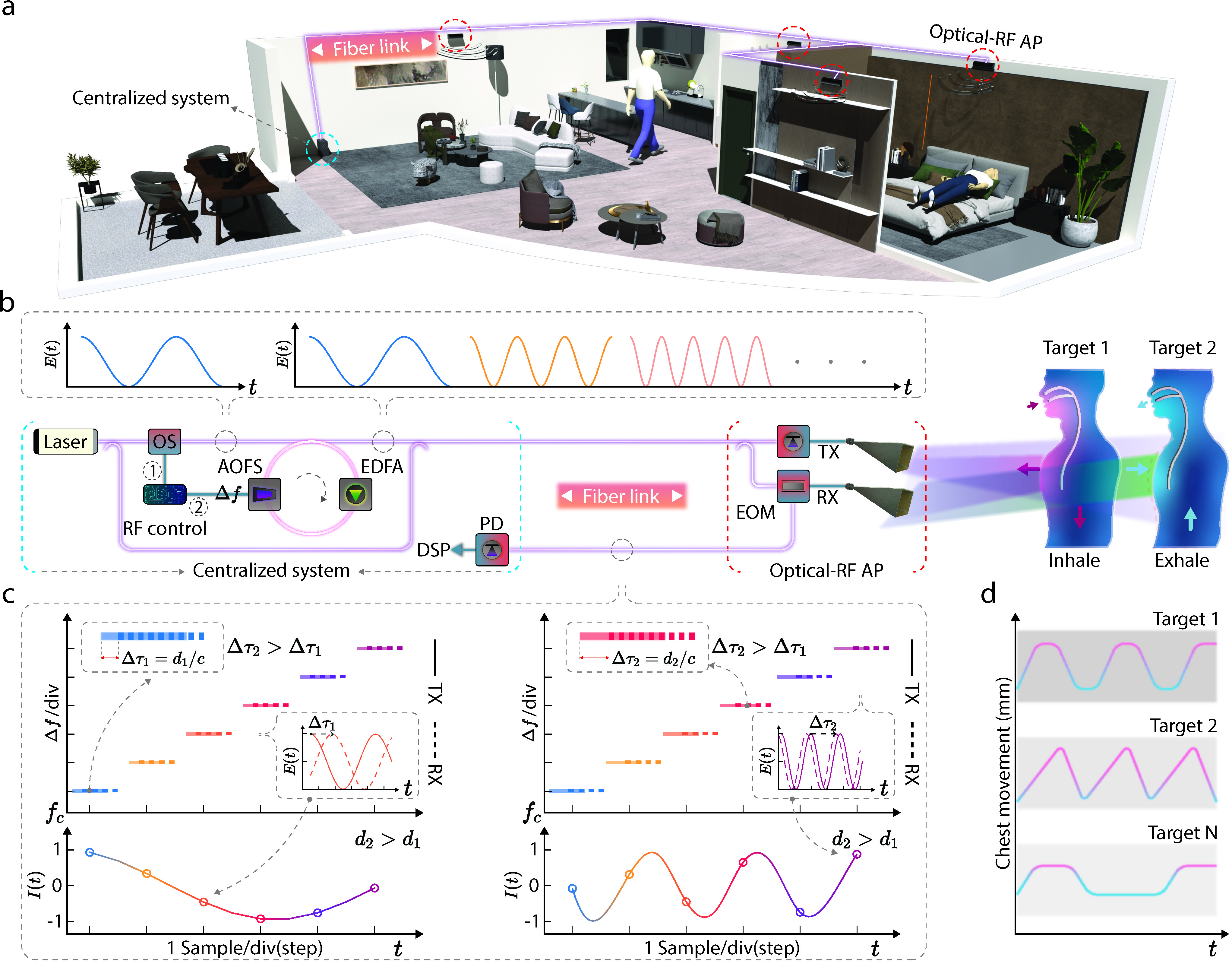}
            }
  \caption{\textbf{A photonics-enabled radar system for contactless vital sign detection.} (\textbf{a}) Conceptual drawings of vital sign radar with distributed sensing access points (APs) enabled by low-loss fiber and a centralized, photonics-assisted radar platform. (\textbf{b}) Schematic of the demonstrated photonic radar for vital sign detection based on a frequency-shifting (FS) fiber cavity. (\textbf{c}) The optical coherent ranging principle is realized by mixing the transmitted and received optical SF signals in a photodetector (PD). After the PD, the demodulated RF signals are illustrated at the bottom panel, demonstrating the ranging at two distance instances, $d_{1}$ and $d_{2}$, respectively. (\textbf{d}) An illustration of using the demonstrated photonic radar with millimeter resolution for detecting respiratory activities from multiple targets. OS, optical switch; FS, frequency-shifting; AOFS, acousto-optic frequency shifter; EDFA, erbium-doped fiber amplifier; EOM, electro-optic modulator; PD, photodetector; DSP, digital signal processing. $\Delta f$, the FS introduced by AOFS.
  }
  \label{fig1}
\end{figure*}
Radar using radio-frequency (RF) waves can remotely access targets' vital signs to overcome the drawbacks of contact-based sensors. Vital sign information is produced based on RF sensing rather than camera filming, naturally providing the desired privacy protection. Electronic radar vital sign detection has recently been explored using single-tone and frequency-modulated (FM) waves. Single-tone radars that rely on the Doppler principle can acquire vital signs by obtaining the phase information of the reflected signal from a moving object. However, this technique lacks the basis to detect the round-trip time to access targets' range information. As a result, they cannot utilize range information to separate closely-located targets and isolate the target from surrounding clutter \cite{Li2013a, Tu2016}, which limits the performance and practicality in real-world deployments. In contrast, FM radars can extract the range information to overcome this issue \cite{Wang2014, Quaiyum2017, Ahmad2018, Mercuri2019}. More importantly, the range resolution and accuracy of FM radars can be increased by broadening the sensing signals' bandwidth. However, conventional electronic radar systems usually have limited sub-GHz bandwidths that lead to tens of centimeters resolution \cite{Wang2013a, Su2019, Fang2020}, which is insufficient to accurately detect delicate human vital sign signals (e.g., human respiration with chest displacement of around 1 cm). This limited resolution would greatly limit the capability of adequately canceling body motions and tracking multiple targets. Moreover, distributed sensing at multiple frequency bands and deployment locations is needed in emerging applications. However, it is challenging for conventional electronics without a complicated parallel hardware architecture \cite{Serafino2021}.

\begin{figure*}[ht!]
 \centering{
  \includegraphics{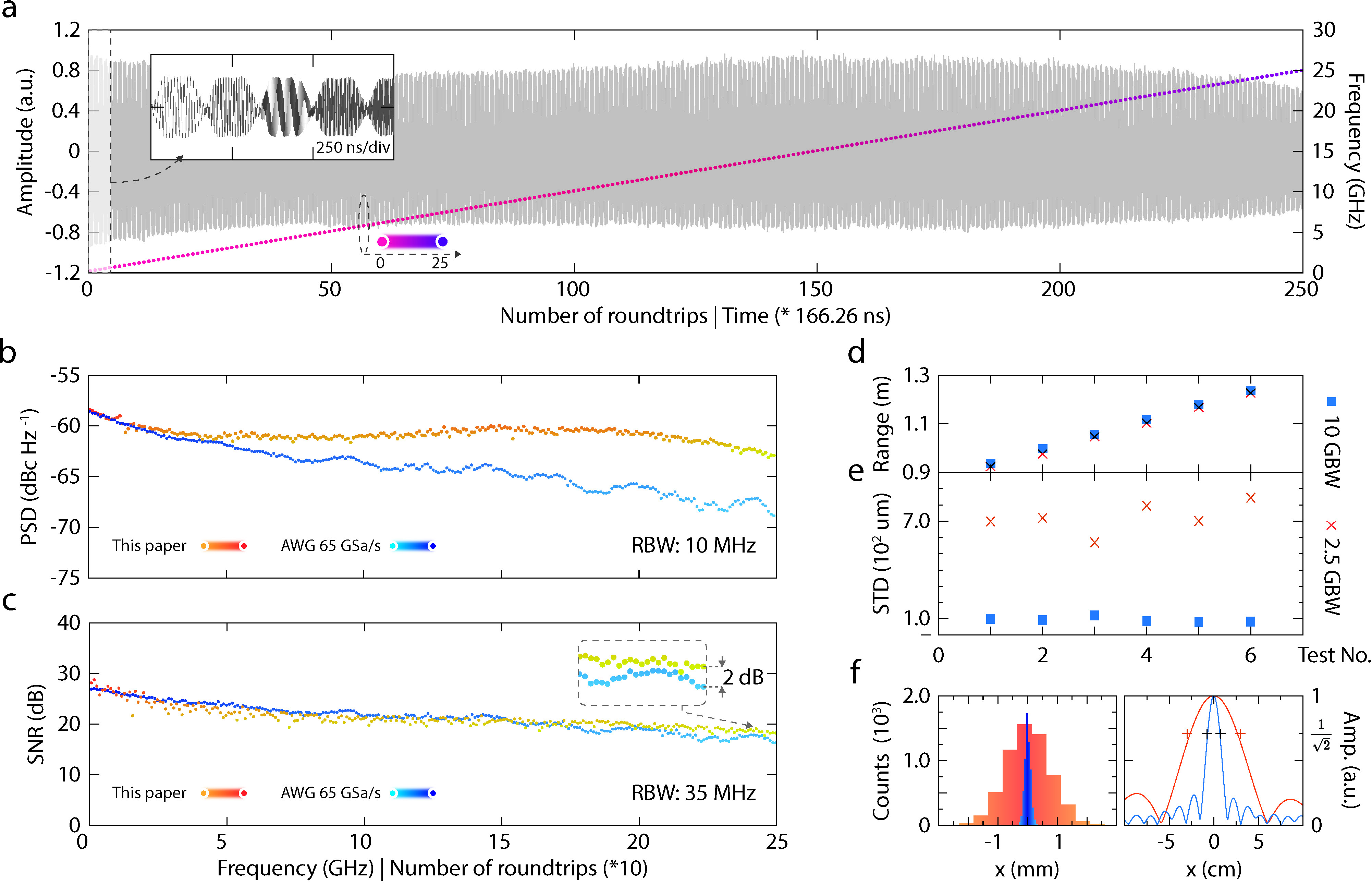}
  }
  \caption{\textbf{Radar signal and ranging quality analyses.} (\textbf{a}) The demonstrated SF waveform with the corresponding frequency shows an overall bandwidth of 25 GHz. (\textbf{b}) The demonstrated signal's power spectrum density (PSD) is compared with an SF signal generated from an arbitrary waveform generator (AWG) with a sampling rate of 65 GSaps. (\textbf{c}) Signal-to-noise ratio (SNR) of the signals generated by the demonstrated system and the AWG. (\textbf{d, e}) Ranging and accuracy, in terms of standard deviation (STD), results using 2.5 GHz (cross) and 10 GHz (square) bandwidth signals. (\textbf{f}) STDs (left) of the 2.5 GHz and 10 GHz SF signals. Over 6,000 measurements, the 2.5 GHz and 10 GHz SF signals show an STD of 725.90 $\mu$m and 93.28 $\mu$m, respectively. Experimental and theoretical (-3 dB, plus) ranging resolutions of the two signals are plotted accordingly.
 } 
\label{fig2}
\end{figure*}
Photonics-assisted approaches have shown significant advantages in achieving wideband and high-resolution radars \cite{Pan2020, Hao2020a} with the capability to generate different formats of radar signals, such as the linear-frequency modulated (LFM) \cite{Peng2018, Li2014a, Cheng2019a, Tang2020} and stepped-frequency (SF) signals \cite{Zhang2020a, Zhang2021, Lpr2022, Zhang2022, Lyu2022}. They are flexible to operate at multiple frequency bands \cite{Serafino2019, Falconi2021, Zhang2018, Xiao2021, Liu2020} across the millimeter-wave region, adapting the radar carrier frequencies for performance optimization based on operating conditions (e.g., weather and target material \cite{Ghelfi2016}). These attractive features overcome the limitations of their electronic counterparts, making them well-suited for vital sign detection. Moreover, photonic radar allows for the potential to achieve a simplified, centralized system using low-loss fiber-based radar signal distribution without scaling the number of electronic components. However, photonic radar systems for vital sign detection have so far remained largely unexplored in real-world scenarios.

Here, we demonstrate a photonic radar for vital sign detection using human respiration simulators and a living animal -- a cane toad -- serving as a human proxy. This radar optical generates 10-GHz-wide stepped-frequency (SF) RF signals in the Ka-band (26.5-40 GHz) to detect respiratory activities of the simulators, achieving 13.7 mm range resolution with a $\mu$m-level accuracy. Such high resolution and accuracy are essential to resolve the delicate vital signs of the cane toad, even with an undersized animal radar cross-section. We demonstrated the bandwidth scalability up to 30 GHz without the limitation of the RF antennas and amplifiers. We further demonstrated a LiDAR vital sign detection system based on the same microwave photonic source, showing the system's potential to enable complementary features of radar and LiDAR. We envisage applying such a high-performance, distributed radar system in various healthcare scenarios, such as round-the-clock vital sign monitoring in aged care facilities, hospitals, and custodial settings. For instance, a distributed photonic radar sensing network with multiple radar optical-RF access points (APs) uses RF waves to detect human vital signs (\fref{fig1}\textbf{a}). This approach can continuously track uncooperative, back- or side-facing targets compared with only a single radar AP deployment. As illustrated, one optical radar signal source enabling multiple optical-RF APs could cover diverse perspectives to monitor one or multiple targets using low-loss optical fibers.

An advanced SF photonic radar is employed to detect the respiratory activities of humans and animals (\fref{fig1}\textbf{b}). The system is mainly structured with an optical frequency-shifting (FS) fiber cavity to generate radar signals in the optical domain, an optical fiber distribution network, and optical-RF APs for electro-optic conversion and RF transceiving. In the FS fiber cavity, an acousto-optic frequency shifter (AOFS) shifts the frequency of an optically injected pulse (with a single frequency of $f_{c}$) by 100 MHz in the succession of each round-trip ($ \Delta f = 100$ MHz). An erbium-doped fiber amplifier (EDFA) inserted in the optical loop compensates for the power loss from optical propagation and coupling. This approach generates an optical SF signal consisting of a series of sine waves with linearly increased frequency at a precisely determined step (\fref{fig1}\textbf{b} and \fref{fig1}\textbf{c}). In the optical-RF APs, SF radar signals in the RF domain are generated through heterodyne mixing with a reference laser (see Methods). The optical signal source only requires cost-effective and low-speed electronic devices (a dual-channel electrical function generator with an analog bandwidth of 100 MHz) to precisely control the total synthesized bandwidth of the SF signal, tunable from sub-GHz level to 30 GHz (see Methods). The function generator could be replaced using 100 MHz reference oscillators and RF switches with reduced complexity.

\begin{figure*}[ht!]
 \centering{
  \includegraphics{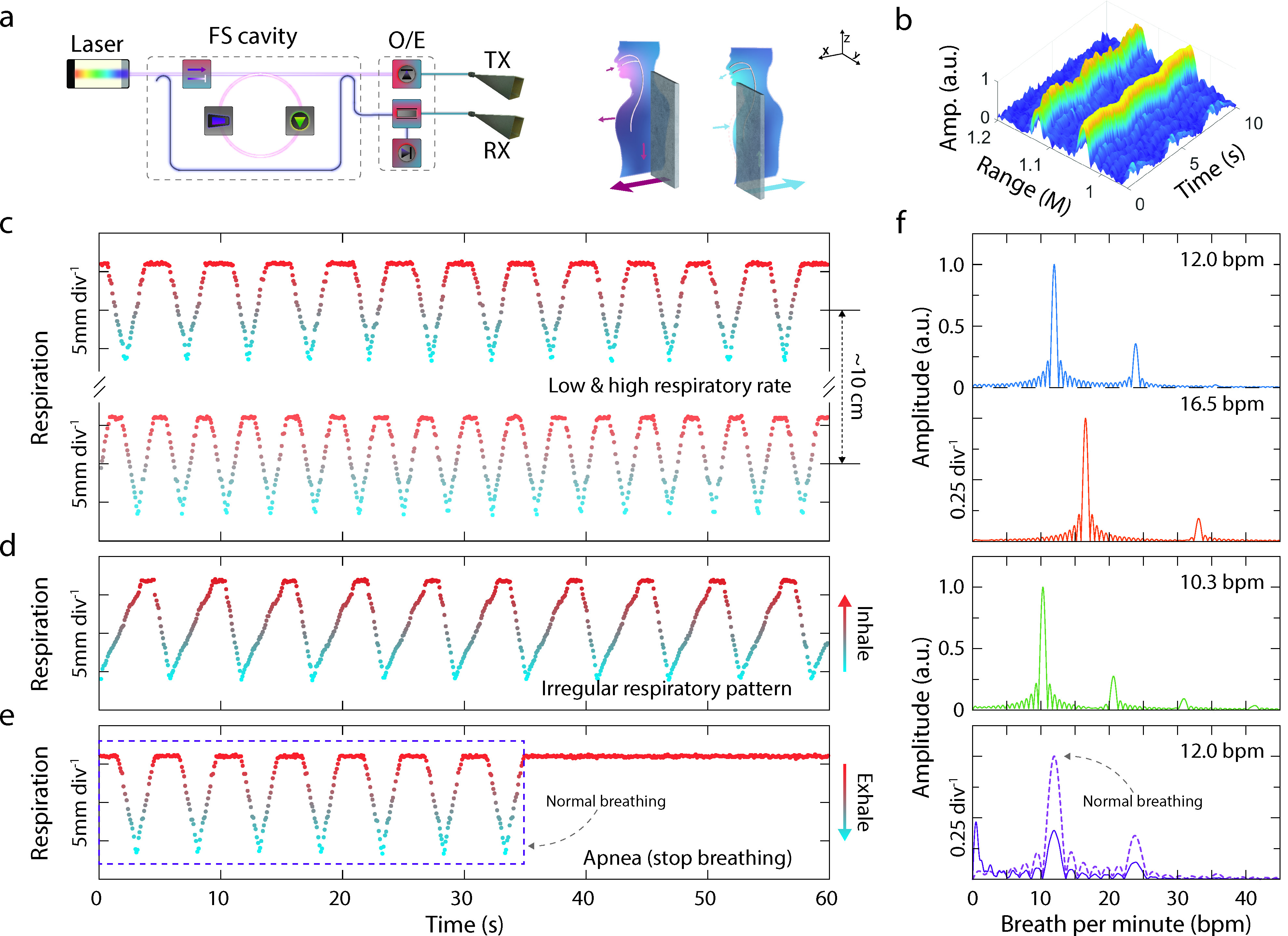}
  }
  \caption{\textbf{Multi-target vital sign detection results based on respiration simulators.} (\textbf{a})  Experimental setup using the demonstrated photonic SF radar with an RF bandwidth of 10 GHz. Two metal plates ($4\times5\times0.3$ cm) separately mounted onto two stepper motors are used here to emulate the chest movement of human breathing. (\textbf{b}) Contactless vital sign detections of two closely located targets. (\textbf{c}) Chest movements of the two targets. (\textbf{d}) Chest movement showing irregular (longer inhale and shorter exhale) breathing patterns. (\textbf{e}) Chest movement shows stopped breathing, which could be a sign of apnea. (\textbf{f}) Fourier transform results based on the chest movement in Fig. 3\textbf{c-e}.
  }  
 \label{fig3}
 \end{figure*}
The radar vital sign detection starts with using the SF signal in the RF domain to illuminate the targets' area of interest (e.g., the chest area for humans) using a transmitting antenna element (TX). Another antenna element (RX) receives the reflected radar signals that carry the vital sign information, which is converted back to the optical domain using an electro-optic modulator (EOM). Demodulated SF signals are generated through a coherent detection process (\fref{fig1}\textbf{c}), i.e., optically mixing the transmitted reference signal (solid lines) with the received signal (dashed lines). Thus, targets at different ranges, e.g., $d_{1}$ and $d_{2}$ have demodulated signals with different oscillating frequencies (\fref{fig1}\textbf{c}). These oscillating frequencies can be extracted through Fourier analysis, showing different peak locations on the frequency domain \cite{Lpr2022}. One advantage of using the SF signal format over other FM approaches is that the radar receiver has a much lower sampling rate favored for fast signal processing, owing to the fact that only one sample is required per round-trip time. For the same bandwidth, the SF signals sustain the exact resolution as other wideband radar waveforms, e.g., the LFM \cite{Ozdemir2021}. As a result, such a system requires less computational power for digital signal processing, enabling real-time, multi-target respiration detection (\fref{fig1}\textbf{d}).

In a vital sign detection radar system, RF sensing signals with broader bandwidth and high time-frequency linearity \cite{Ayhan2016} are required to sustain the range resolution and reduce measuring errors (increase accuracy). The signal (\fref{fig2}\textbf{a}) generated by the FS cavity demonstrated a 25 GHz bandwidth (30 GHz in Methods) to compare with the 25-GHz analog bandwidth of an arbitrary waveform generator (AWG). The FS cavity has a round-trip time of 162.26 ns with a constant 100-MHz FS enabled by the AOFS. Thus, a total synthesized 25 GHz SF signal can be generated after the first injected pulse finishing 250-time round-trips in the FS cavity (see Methods). The high time-frequency linearity is inherited from the constant and stable AOFS-enabled frequency-time shifting \cite{Zhang2022}, which is challenging to achieve using alternative photonics-based approaches \cite{Hao2018, Zhou2022a}.

\begin{figure*}[ht!]
 \centering{
  \includegraphics{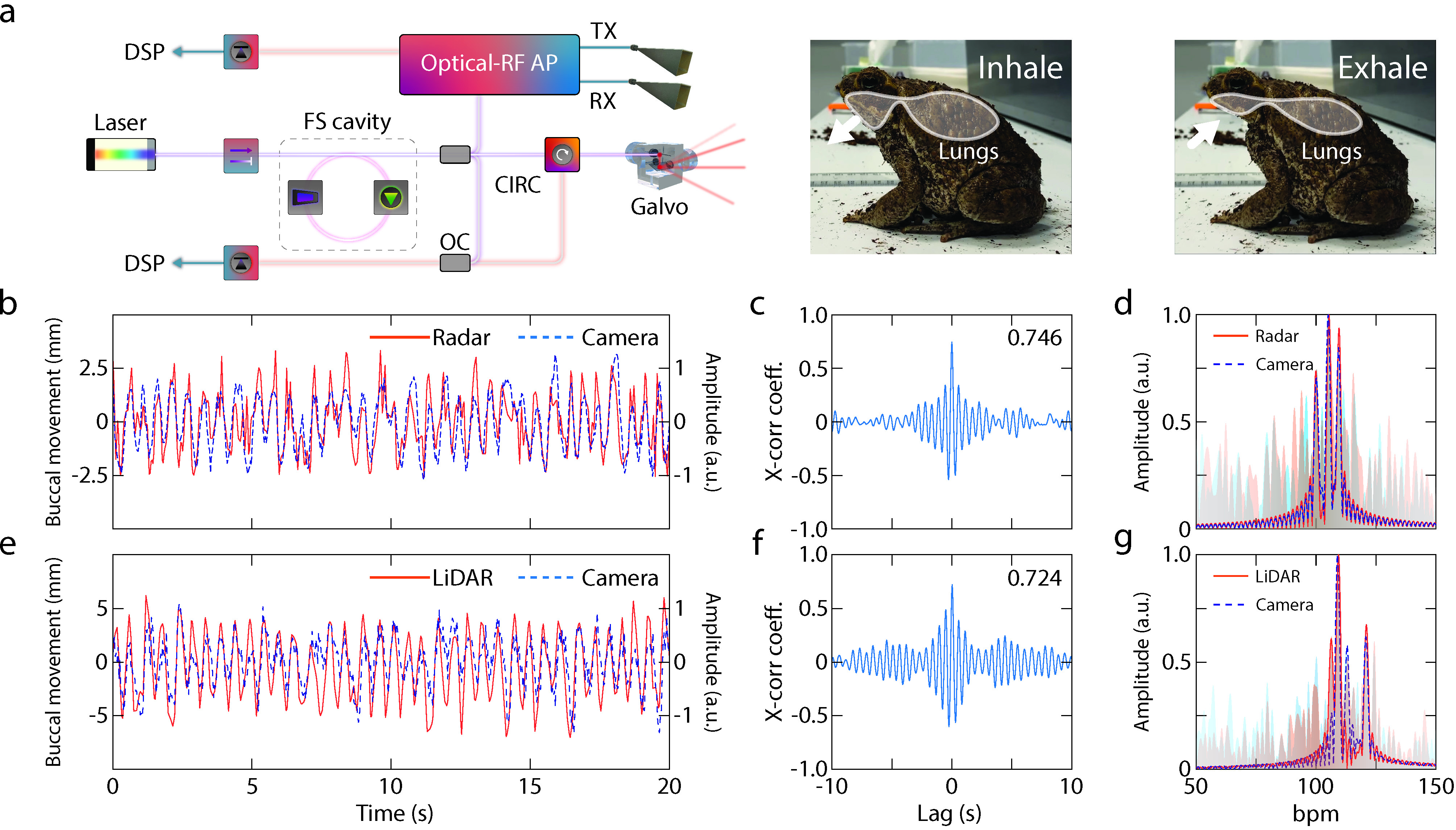}
  }
  \caption{\textbf{Vital sign detection results using a cane toad as a human proxy.} (\textbf{a}) Experimental setup using the demonstrated system for radar sensing, which also shows its flexibility and capability enabling LiDAR detection. (\textbf{b}) Experimental results of detecting the buccal cavity movement of the cane toad using the demonstrated radar system. Camera-extracted data is used as a reference. (\textbf{c}) Cross-correlation between the radar and camera data, showing a coefficient of 0.746. (\textbf{d}) Frequency domain analyses of the radar (red shadow) and camera data (blue shadow) with three top-weighted Fourier coefficients in solid and dash lines, respectively. (\textbf{e}) Experimental results of detecting the buccal cavity movement of the cane toad using a LiDAR system. (\textbf{f}) Cross-correlation between the LiDAR and camera data, showing a coefficient of 0.724. (\textbf{g})  Frequency domain analyses of the LiDAR (red shadow) and camera data (blue shadow) with three top-weighted Fourier coefficients in solid and dash lines, respectively. OC: optical coupler; CIRC: optical circulator; Galvo: 2-D scanning mirrors.
  }  
  \label{fig4}
\end{figure*}
The power spectral density (PSD) of the generated SF waveform sustains a constant level of around -60.0 dBc/Hz during the entire recirculation (\fref{fig2}\textbf{b}) and slightly reduces to -62.4 dBc/Hz when reaching the 250-time circulation. In contrast, the PSD of the electronic counterpart exhibits noticeable degradation, from - 57.96 dBc/Hz to -68.30 dBc/Hz, due to the AWG's limited adequate analog bandwidth. The signal-to-noise ratio (SNR) of the generated is also calculated based on the PSD results (see Methods) and cross-examined with the signal generated by the AWG (\fref{fig2}\textbf{c}). The SNR results show no significant difference between these two signals across the 25 GHz bandwidth, proving that the demonstrated system could generate radar signals with AWG-comparable quality. It is worth mentioning that the signal quality is not expected to deteriorate through photonics-based up-conversion as opposed to the conventional electronic up-conversion \cite{Ho2010, Jankiraman2018a}.  

Furthermore, we investigate the range resolution (\fref{fig2}\textbf{d}) and accuracy characterized in standard deviation (STD) (\fref{fig2}\textbf{e}) of the presented vital sign detection photonic radar system. The results were measured using 2.5 GHz (cross) and 10 GHz (square) bandwidth SF signals based on a metal plane reflector with a dimension of $4\times5\times0.3$ cm (see Methods). The SF signal with a wider bandwidth (10 GHz) shows a significant accuracy improvement compared with the narrower band (2.5 GHz) SF signal (\fref{fig2}\textbf{f}), revealing a reduction in STD to 93.28 $\mu$m from 725.90 $\mu$m based on the signals with experimental range resolutions of 13.7 and 53.2 mm, respectively. These results proved that increasing the sensing signal bandwidth will simultaneously improve the range resolution and accuracy, which is greatly preferable for a radar system to detect delicate respiratory activities from multiple targets. 

Next, we applied the photonic radar to multi-target respiration detection based on two human breathing simulators (\fref{fig3}\textbf{a}). Our radar can radar successfully detect the respiratory activities from the two closely located targets ($\sim $10 cm apart) in real-time (\fref{fig3}\textbf{b}). Over a 60-second time window, the relative 'chest' movements of the two targets are extracted (\fref{fig3}\textbf{c}). The corresponding respiratory frequencies, in terms of breath per minute (bpm), are acquired by taking the Fourier transform of the trajectories (\fref{fig3}\textbf{f}) with a respiratory rate (RR) of 12 bpm and 16.5 bpm for the top and bottom target, respectively. These two frequencies are chosen deliberately to fit in the typical RR of an adult at rest \cite{Chioukh2011, Schriger2020, Nahar2018}. Meanwhile, the superior accuracy allows the radar system to detect irregular respiration patterns with subtle movements (around mm-level), such as irregular breathing with longer inhales and shorter exhales and stop breathing (\fref{fig3}\textbf{d} and \fref{fig3}\textbf{e}). Therefore, it could help accurately identify or even predict respiratory abnormalities linked to many medical conditions such as asthma, anxiety, congestive heart failure, and lung disease \cite{Schriger2020}. The demonstrated results proved that the accuracy enabled by the demonstrated system offers sufficient precision to pick up respiratory abnormalities.     

To prove its suitability in practical applications, we demonstrate animal respiratory detection using a female cane toad as a human proxy (a pilot study before human trials) to evaluate the radar performance (\fref{fig4}\textbf{a}). The cane toad has a radar cross-section (the buccal area, $\sim  2 \times 2.5$ cm) smaller than the human chest, making the experiment more challenging than human trials. The buccal cavity was connected to the lungs as a part of its air-breathing activity \cite{Stinner2000, Reid2000, Jenkin2011}. The toad was located about 1 meter from the radar antenna, with the beam pointing to the toad's buccal area. The data extracted from the photonic radar reveals the real-time trace of the toad's buccal movement with a displacement of around 5 mm (\fref{fig4}\textbf{b}). Meanwhile, the radar data is cross-referenced with the data extracted from a video clip (see Methods) recorded simultaneously, showing a cross-correlation coefficient of 0.746 (\fref{fig4}\textbf{c}). It is worth mentioning that the camera has a different perspective from the radar beam direction, which might slightly decrease the correlation coefficient. The Fourier domain analyses based on radar and camera data further prove the accuracy and performance of the photonics-enabled radar system (\fref{fig4}\textbf{d}). The respiration data shows that the cane toad has an irregular respiration pattern, owing to the fact that intermittent or discontinuous breathing patterns (see Cane Toad Respiration) are common in amphibians \cite{Stinner2000, Reid2000, Jenkin2011}, which agrees with both the radar and camera-based results.

Attractively, the proposed radar system can simultaneously be adopted for LiDAR sensing (\fref{fig4}\textbf{a}), of which the RF components, such as the RF amplifier and antennas, no longer limit the system's bandwidth. The demonstrated system enables the LiDAR system with a total synthesized bandwidth of 25 GHz (6 mm range resolution), sufficient to catch the toad's buccal movement (\fref{fig4}\textbf{e}). The performance is also validated by cross-referencing the LiDAR data with the camera data using cross-correlation (\fref{fig4}\textbf{f}) and Fourier analyses (\fref{fig4}\textbf{g}). This provides an approach to achieve a hybrid radar-LiDAR system that can combine complementary detection techniques for improved sensing accuracy and system resilience.

In summary, we have demonstrated a photonic vital sign detection radar system with a fine resolution down to 6 mm and $\mu$m-level accuracy, enabling multi-target detection without comfort and privacy issues. Experimental validations proved its ability and effectiveness in picking up delicate respiratory abnormalities and accurately extracting the cane toad's buccal movement. More importantly, it has a simplified system structure with improved bandwidth and flexibility that the current state-of-the-art electronic vital sign radars cannot achieve without requiring parallel or multiplexed electronic architectures.

The system can support radar and LiDAR sensing, proving its unprecedented flexibility and potential for hybrid detection and sensor fusion that offers more consistent and accurate sensing results \cite{Falconi2021}. On the one hand, radar signals can penetrate clothes and thin tissues to provide vital signs data such as respiration and heart rate. On the other hand, the LiDAR system has significantly improved spatial resolution and directionality to provide superior information on the targets' body motion and the surroundings, serving as a complementary approach to assist and optimize the vital sign radar's performance without compromising privacy.

The demonstrated system is compatible with the photonic distributed, multi-band operation radar technique \cite{Serafino2021}. Thus, multiple sensors, enabled by one centralized photonic system, can seamlessly work together without interference for a broader detecting coverage with lower overall complexity and cost. For future development towards the miniaturization of the photonic system, recent advances in photonic integration of critical function building blocks, such as on-chip acousto-optic frequency shifter \cite{Shao2020, Yu2021} and optical waveguide amplifiers \cite{Vagionas2022, Liu2022}, provide a promising technical basis to achieve a compact size for portable sensing \cite{Falconi2021, Li2020}. This novel photonic approach offers a new path toward high-resolution, rapid-response, and cost-effective hybrid radar-LiDAR modules for distributed, contactless vital sign detection.

\vspace{15pt}
\small
\noindent \textbf{Acknowledgments} This work was supported in part by the U.S. Air Force (USAF) under Grant FA2386-16-14036, in part by the U.S. Office of Naval Research Global (ONRG) under Grant N62909-18-1-2013, and in part by Australian Research Council Discovery Project under Grant DP200101893. We acknowledge the discussion on contactless vital sign detection with A. Withana and the assistance in building the respiration simulator with J. Coyte. Z.Z. is supported by an Australian Government Research Training Program (RTP) Scholarship.

\noindent \textbf{Data Availability} The data supporting this study's findings are available upon request.

\noindent \textbf{Animal Ethics} All animal experiments were performed under the Animal Research Authority, project number 2022/2090, approved by the Animal Ethics Committee (AEC), the University of Sydney, in compliance with Section 27 of the NSW \textit{Animal Research Act 1985}.

\noindent \textbf{Author Contributions} Y.L. and B.J.E. conceived the project; Z.Z. and Y.L. designed the system structure; Z.Z. conducted the experiments; Z.Z. and T.S. conducted the animal experiments; Z.Z., Y.L., and T.S. performed the data analysis; Z.Z. and Y.L. wrote the manuscript with contributions from B.J.E. and T.S.; Y.L. and B.J.E. supervised the project.

\bibliography{ms.bib}
\end{document}


\title{Photonic Radar for Contactless Vital Sign Detection: Supplemental Document}
\author{Ziqian Zhang}
\affiliation{Institute of Photonics and Optical Science (IPOS), School of Physics, The University of Sydney, NSW 2006, Australia}
\affiliation{The University of Sydney Nano Institute (Sydney Nano), The University of Sydney, NSW 2006, Australia}
\author{Yang Liu}
\email{yang.liu@sydney.edu.au}
\affiliation{Institute of Photonics and Optical Science (IPOS), School of Physics, The University of Sydney, NSW 2006, Australia}
\affiliation{The University of Sydney Nano Institute (Sydney Nano), The University of Sydney, NSW 2006, Australia}
\author{Tegan Stephens}
\affiliation{Bird and Exotics Veterinarian, NSW 2017, Australia}
\author{Benjamin J. Eggleton}
\email{benjamin.eggleton@sydney.edu.au}
\affiliation{Institute of Photonics and Optical Science (IPOS), School of Physics, The University of Sydney, NSW 2006, Australia}
\affiliation{The University of Sydney Nano Institute (Sydney Nano), The University of Sydney, NSW 2006, Australia}
\date{\today}
\maketitle

\section*{Methods}
\begin{figure*}[ht!]
  \centering{
  \includegraphics{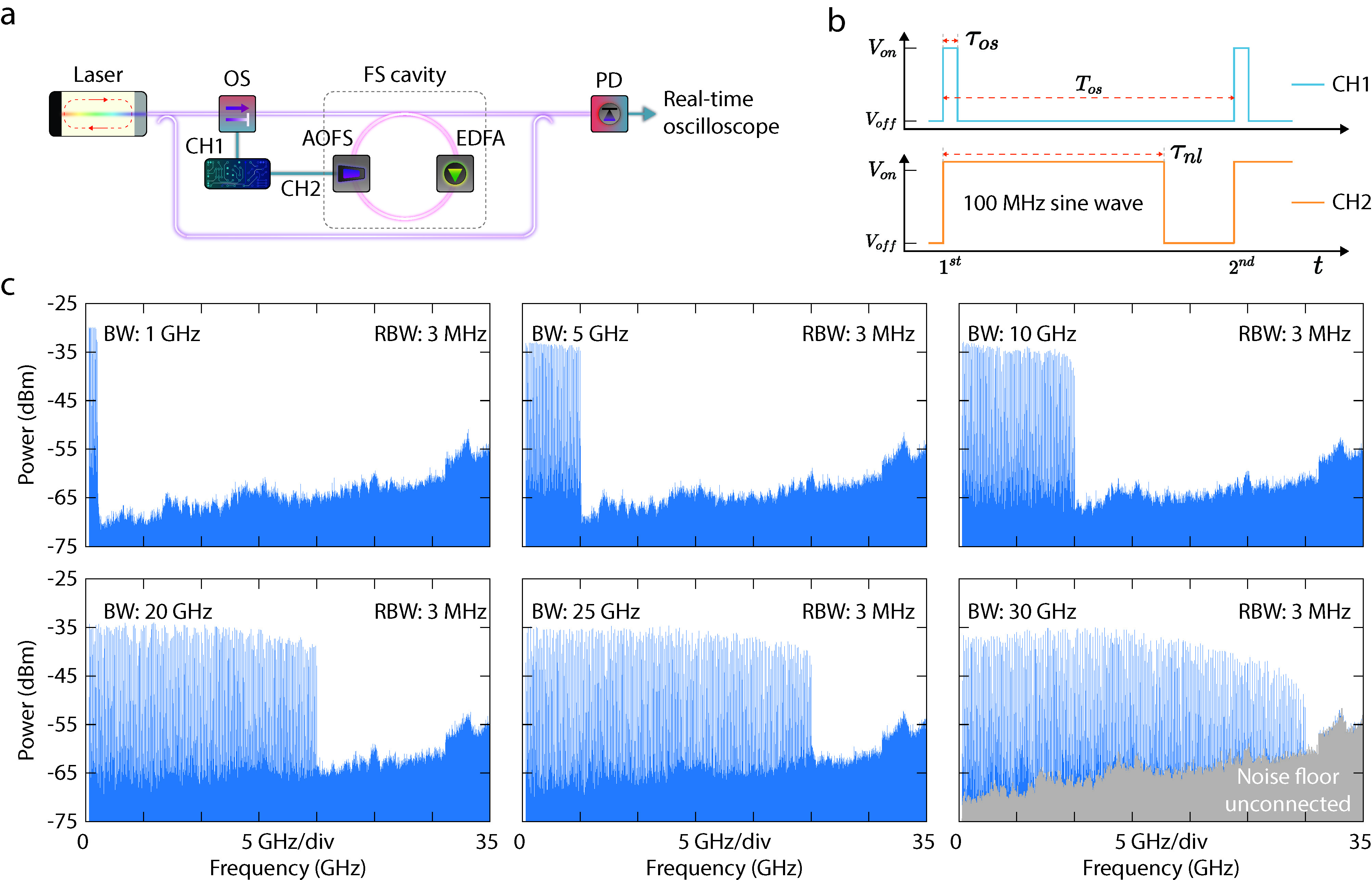}
  }
  \caption{\textbf{Flexible bandwidth tuning based on a software-definable MHz-level RF source.} (\textbf{a}) System schematic for generating based SF signals in the RF domain. (\textbf{b}) MHz-level system control signals generated by a two-channel RF source. (\textbf{c}) Baseband RF bandwidth tuning from 1 to 30 GHz by adjusting the RF control signals.}
\label{FS6}
\end{figure*}

A laser is connected to an optical coupler as the optical source in the system (\fref{FS6}\textbf{a}). A dual-channel function generator generates the RF control signals that drive the FS cavity with an analog bandwidth of 100 MHz. As illustrated in \Fref{FS6}\textbf{b}, one channel (CH1) generates a periodic on-off signal to control an optical switch (OS) so that the optical pulse has a dwell time of $\tau_{os}$ and a repetition rate of $T_{os}$. The other channel (CH2) generates a $\Delta f$-Hz sine wave to drive the AOFS. In order to control the total number of roundtrips $N$ (the synthesized bandwidth), a rectangular envelope is applied on top of the $\Delta f$-Hz sine wave in CH2 (\fref{FS6}\textbf{b}), with a duty cycle of $\tau_{nl}$ ($N\times \tau_{os}$) and repetition rate of $T_{nl}$ to switch on-off the AOFS \cite{Lyu2022}. Thus,  bandwidth tuning from the sub-GHz level to 30 GHz can be realized by adjusting the duty cycle $\tau_{nl}$ (\fref{FS6}\textbf{c}).

\begin{figure*}[ht!]
  \centering{
  \includegraphics{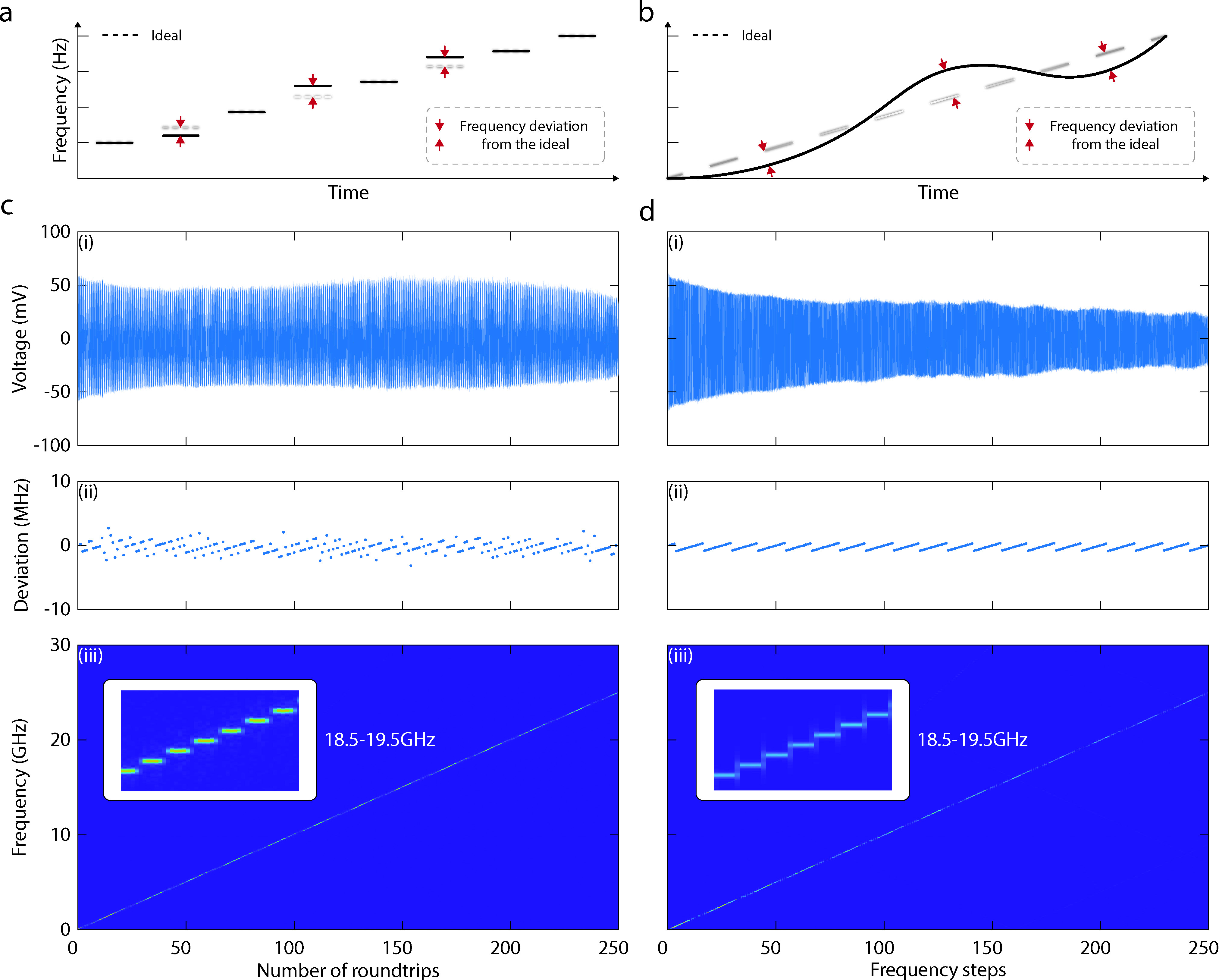}
  }
  \caption{\textbf{Demonstration of the signal generated by the FS cavity has linearity comparable to a high-speed signal generator in terms of the root mean square (RMS).} (\textbf{a}) Illustration of non-linearity existed in SF signals. (\textbf{b}) Illustration of non-linearity existed in LFM signals. (\textbf{c}) The time-domain, frequency deviation, and time-frequency plots of the SF signal generated based on the FS cavity. The RMS in (ii) is 955 kHz. (\textbf{c}) The time-domain plot, frequency deviation, and time-frequency plot of the SF signal generated using a signal generator with a speed of 65 GSa/s. The RMS in (ii) is 435 kHz.}
\label{FS1}
\end{figure*}

Frequency-modulate (FM) signals with high time-frequency linearity are essential for accurate radar and LiDAR ranging. FM signals, e.g., SF (\fref{FS1}\textbf{a}) and LFM (\fref{FS1}\textbf{b}), with low linearity, exhibit significant deviation between the ideal and real signals. Here, we use these frequency deviations' root mean square (RMS) to evaluate the SF signal's linearity generated by the demonstrated system (\fref{FS1}\textbf{c}) and a signal generator (\fref{FS1}\textbf{d}) with a speed of 65 GSa/s. The demonstrated approach shows a similar linearity performance compared with the state-of-the-art signal generator, revealing an RMS of 955 kHz over a bandwidth of 25 GHz compared to a 435 kHz RMS from the signal generator. 

Baseband RF signals are acquired by mixing the optical SF signal with a tap of the CW laser in a 50 GHz photodetector (PD). A real-time oscilloscope, with a sampling speed of 80 GSa/s, measures the time-domain data in Figure 2. Each point of the time and frequency data is extracted based on a 100 ns time window located at the middle of each frequency step. 

\begin{figure*}[ht!]
  \centering{
  \includegraphics{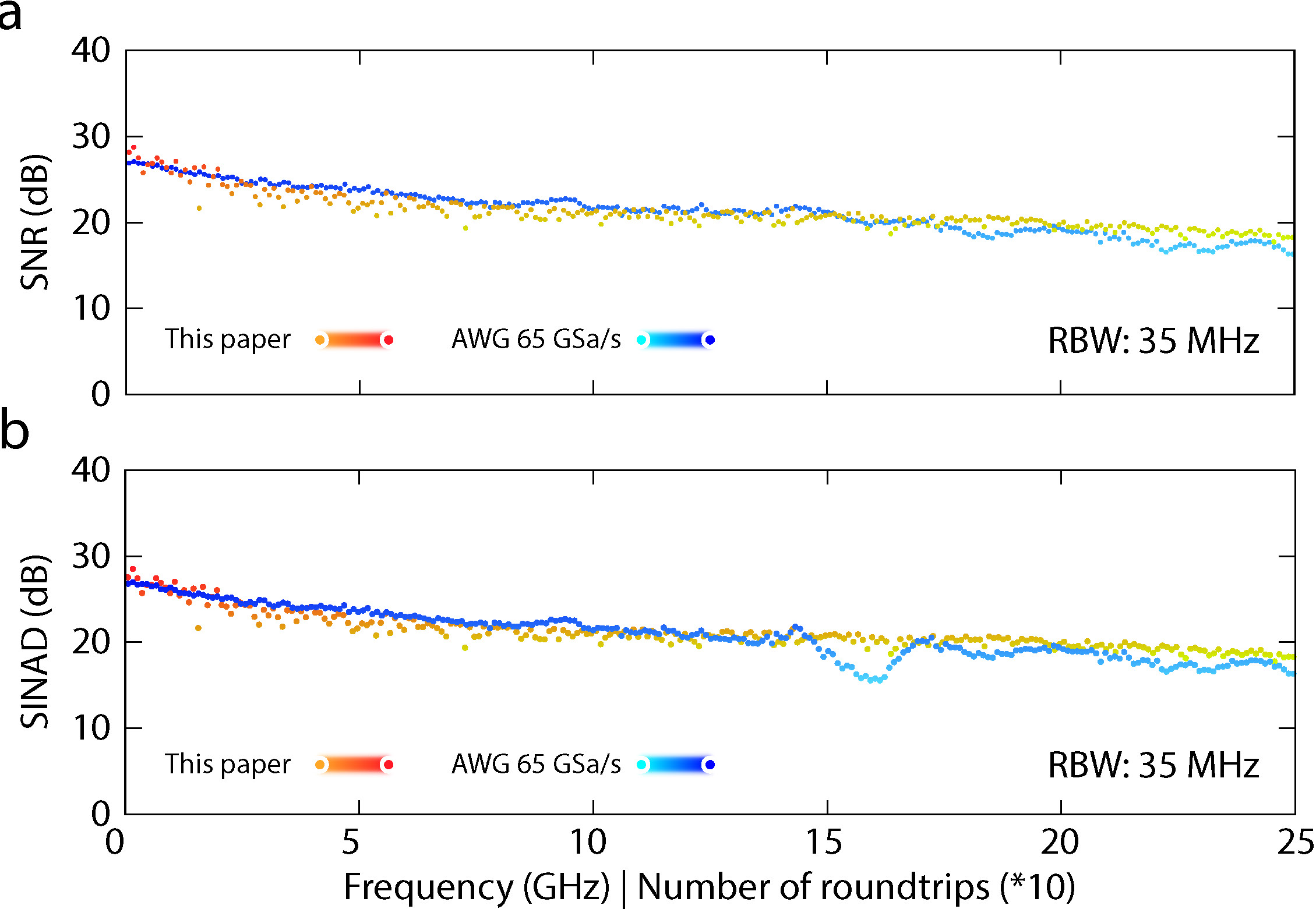}
  }
  \caption{(\textbf{a}) SNR calculation based on PSD. (\textbf{b}) Signal-to-noise and distortion ratio (SINAD) calculation based on PSD.}
\label{FS5}
\end{figure*}

The power spectral density (PSD), shown in Figure 2\textbf{b}, is estimated based on a one-side periodogram \cite{Fulop2006a} using the time-domain data shown in \Fref{FS1}. Each PSD estimation uses a 100 ns time-domain clip with 8000 points rectangular window and a resolution bandwidth of 10 MHz. The signal-to-noise-ratio (SNR), shown in Figure 2\textbf{c}, is estimated based on a one-side power spectrum using a 100 ns time-domain clip with 8000 points Kaiser window (shape factor $\beta = 38$) to maximize the energy concentration in the main lobe \cite{Oppenheim1999}, which has an equivalent rectangular noise bandwidth of 35.08 MHz. We also calculated the signal-to-noise and distortion ratio (SINAD) to evaluate the signal quality further, based on the PSD shown in \Fref{FS5}\textbf{b}. 

The optical radar coherent demodulation is realized by mixing the transmitted and received signal in a PD. An electronic low-pass filter is connected to the RF output of the PD to filter out unwanted high-frequency RF signals. The demodulated radar signal in the RF domain will be in the form of time-dependent oscillation, which can be written as $S_{d}(t,n) = \sum^{N}_{n=1}\text{cos}(2 \pi n \Delta f \Delta \tau)$ \cite{Lpr2022}, where $N$ is the total number of recirculation, $\Delta \tau$ is the time delay between the reference SF signal (a copy of the transmitted SF signal) and the reflected SF signal. The signal's roundtrip time delay can be written as $\Delta \tau = 2d/c$, where $d$ is the range of a target and $c$ is the speed of light.

The ranging results are measured using 2.5 GHz and 10 GHz stepped-frequency waveforms to illuminate a metal plane reflector with a size of $4\times5\times0.3$ cm. The reflected RF signal is converted to the optical domain through a phase modulator and mixed with a tap of the transmitted optical SF signal for optical coherent detection in a photodetector, forming an RF demodulated signal (Fig. 1\textbf{c}). Real-time oscilloscope samples the demodulated signal with a sampling rate greater than the inverse of the round-trip time to ensure a minimum of 1 sample per frequency step. The demodulated signal is used in digital signal processing. For instance, the 10 GHz SF signal has 100 steps; thus, 100 data points are used for extracting the range information. Each ranging result has a Fourier transform (FT) limited resolution of 41 $\mu$m using a $2^{16}$-point (inverse) FT. Each data point in Figure 2(\textbf{d}) is extracted using 1000 time range measurements. The 1000-time range measurement is repeated at different distances with a 5 cm spacing and an approximately 1 mm human placement error.

\begin{figure*}[ht!]
  \centering{
  \includegraphics{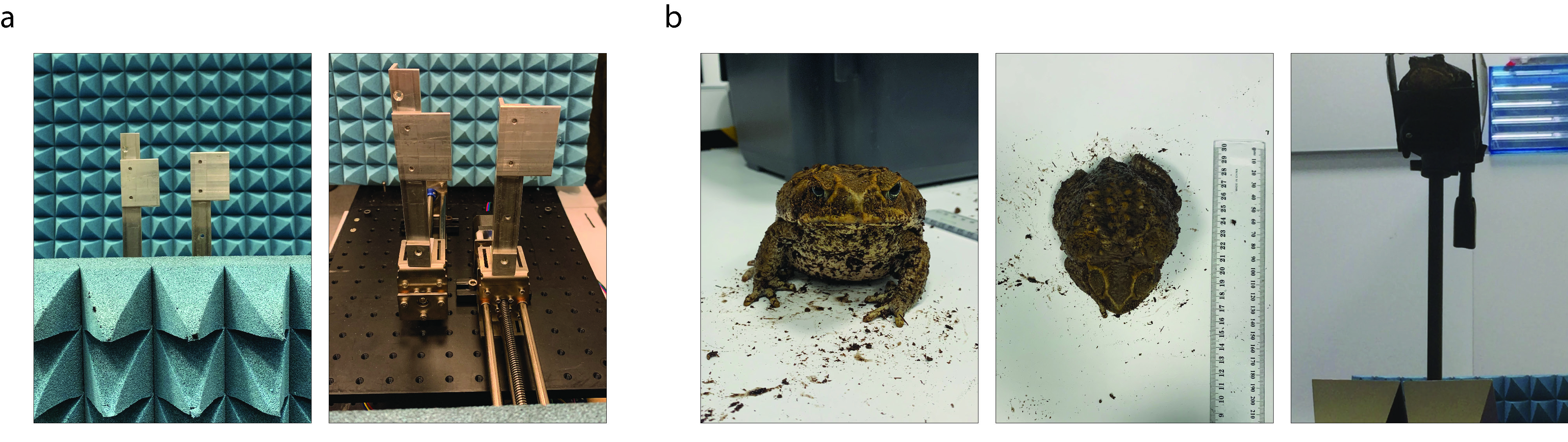}
  }
  \caption{\textbf{Experimental setup of the breathing simulators and the cane toad.} (\textbf{a}) The respiration simulators based on programmable stepped motors and metal plane reflectors with a size of $4\times5\times0.3$ cm. (\textbf{b}) A female cane toad with a body length of approximately 130 mm and the radar detection setup with the antenna's main beam focused at the toad's buccal area.}
\label{FS4}
\end{figure*}

The theoretical radar cross section (RCS) value of the metal plane reflector is 0.503 $m^{2}$ assuming it is a perfect conducting object using a 30 GHz radar signal and the radar is perpendicular to the surface \cite{Chen2014b, Ozdemir2021}. The RCS of the object is chosen in comparison to a human RCS of 1 $m^{2}$ \cite{Rezende2002}. The reflector is also used in the experiment in Figure 3, which is mounted onto a stepped motor to emulate human chest movement while breathing (\fref{FS4}\textbf{a}). It is worth mentioning that the accuracy of the 10 GHz bandwidth signal is expected to be worse than the 93.28 $\mu$m when sensing humans or animals with a smaller radar cross-section.

\section*{Ambiguity range}
The coherent detection can extract the phase differences between the sent and received signals which can be expressed as:
\begin{equation}
    \phi_{n} = 2\pi \cdot n \Delta f \cdot \frac{2d}{c}
    \label{eq2}
\end{equation}where $n = 1,2,...N$, $N \Delta f$ is the total bandwidth, $\Delta f$ is the frequency shift introduced by the AOFS, $d$ is the distance of the target, and $c$ is the propagation speed of the RF signal. To this point, two main approaches can be used to acquire the range information of the target. \Eref{eq2} can also help understand the unambiguous range involved in using SF signals for detection. Consider a relative distance change to the original location $d$ as $\Delta d$, then the corresponding phase changes can be written as \cite{Pinna2017}
\begin{equation}
    \Delta \phi_{n} = 2\pi \cdot n \Delta f \cdot \frac{2\Delta d}{c},
    \label{eq3}
\end{equation} which has a periodical phase change of $2\pi$. If the relative phase changes are more than the period ($2\pi$), ambiguities are introduced ($\Delta \phi > 2\pi$). Thus, by substituting the maximum phase changes, $2\pi$ into \Eref{eq3} and moving $\Delta d$ to the left side, the unambiguous range can be expressed as
\begin{equation}
    \Delta d_{max} = \frac{c}{2 \Delta f}.
    \label{eq4} 
\end{equation}Therefore, the unambiguous range of our system is about 1.5 meters, which can be further extended by reducing the frequency shift $\Delta f$. 

Here are two instances: 1) SF signals with a 1 GHz frequency shift will bring the unambiguous range down to 0.15 meters, and 2) using AOFS with the opposite frequency shifting could reduce the frequency shift to below 10 MHz, which is tantamount to an unambiguous range above 15 meters.
\section*{Camera vital sign detection}
\begin{figure*}[ht!]
  \centering{
  \includegraphics{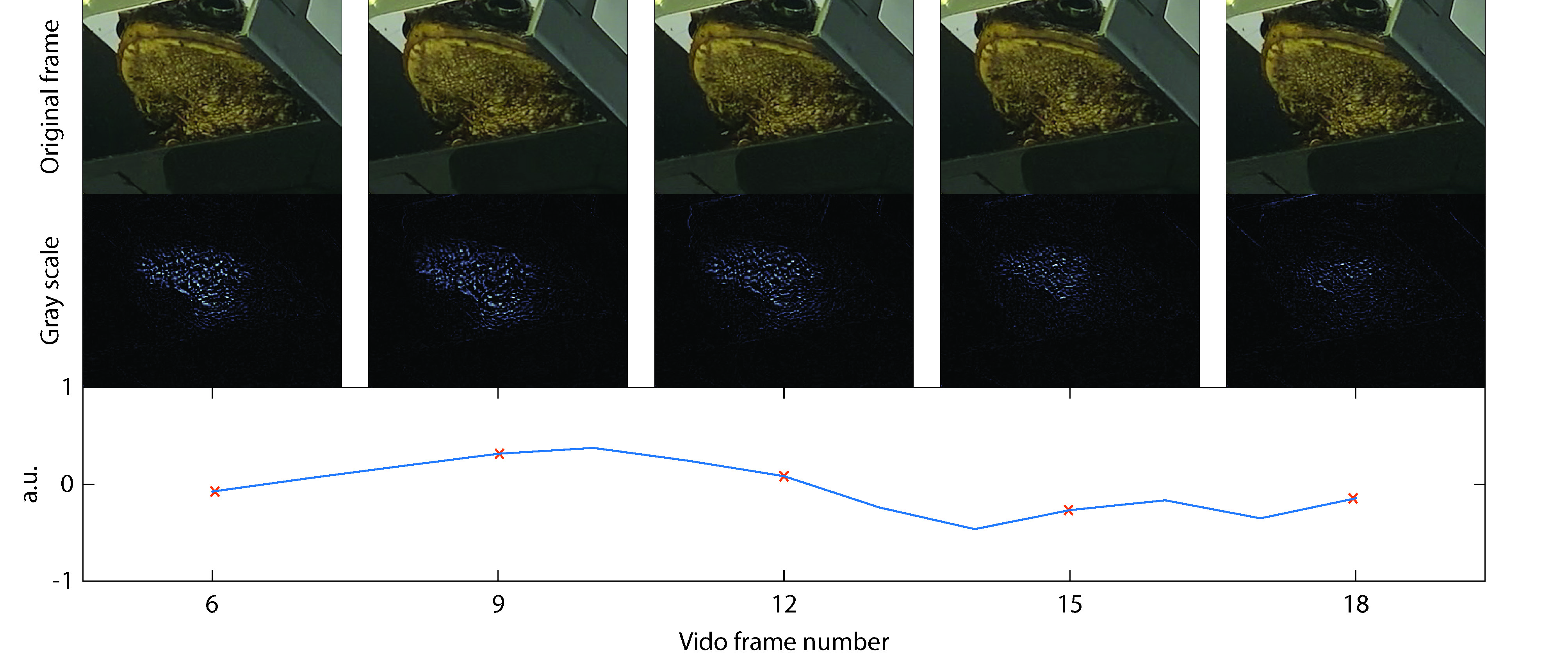}
  }
  \caption{Camera-based respiration extraction of a cane toad.}
\label{FS3}
\end{figure*}

While using the demonstrated photonic radar to detect the buccal movement of the cane toad, we also used a camera to record the activity simultaneously as a reference. The video clip is recorded with a speed of 30 frame-per-second (FPS). Each video frame is converted into a numerical, gray-scale ($[0,1]$), 2D matrix and then subtracted with a reference frame. Therefore, body motions (buccal movement) will be translated into intensity changes, which can be used to extract the cane toad's respiration (\fref{FS3}).
\section*{Cane Toad Respiration}
Cane toads have more complex respiration patterns than humans due to the complexities of multi-organ gas exchange, including skin, gills, and lungs \cite{Jenkin2011}. Lungs are the organ responsible for air-breathing, which contains a series of events causing buccal movement. In general, one air-breathing cycle comprises the following events. First, fresh air is drawn into the lower half of the buccal cavity through buccal depression. Second, the air from the previous air-breathing cycle exits from the mouth (nares) through the upper half of the buccal cavity. Finally, the fresh air is forced into the lungs with two possibilities occurring after this final cycle: 1) the entire cycle is repeated (single breaths), and 2) only the cycle is repeated without lung ventilation (doublets). Either single breaths or doublets have a regular, non-discontinuous interval. Two female cane toads are approved for the experiment. These two cane toads are slightly different in body size. The radar and LiDAR experiments use different cane toads.
\section*{Table for comparison}
\begin{figure*}[ht!]
  \centering{
  \includegraphics{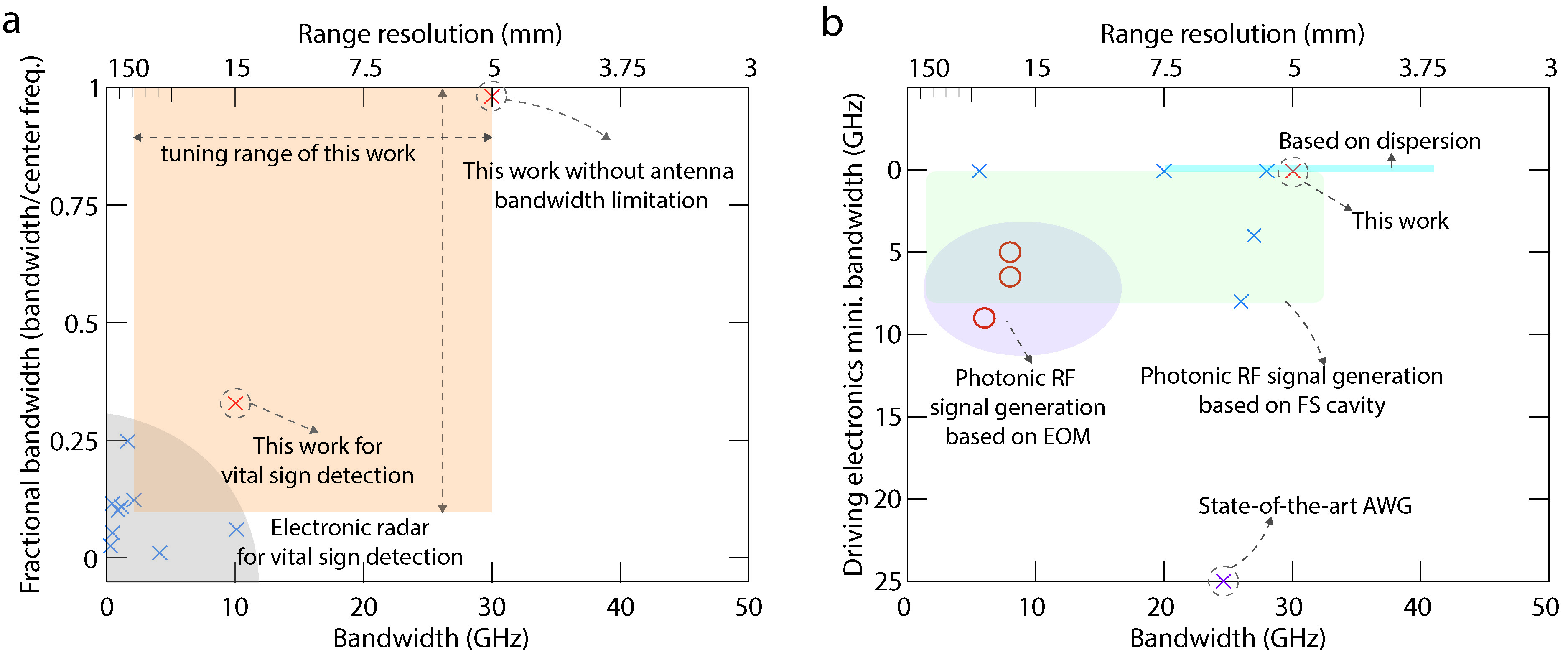}
  }
  \caption{\textbf{Evaluation of the demonstrated photonic radar versus existing electronic vital sign radars and radar signal generation approaches using photonics.} \textbf{(a)} Evaluation of the demonstrated photonic radar system against conventional electronic vital sign radars in terms of fractional bandwidth (the ratio between frequency tuning range and center frequency) and the system's range resolution. \textbf{(b)} Comparison of the photonics-based wideband RF signal generation in terms of the speed of the system's driving electronics and the radar signal bandwidth across various photonic approaches.}
\label{FS7}
\end{figure*}

We provide two tables to compare the demonstrated system with existing electronic vital sign radars (\tref{tab1}) and several photonic approaches for generating radar signals (\tref{tab2}). \Fref{FS7} visualized several figures of merit based on \tref{tab1} and \tref{tab2}. \Fref{FS7}\textbf{a} evaluates the demonstrated photonic radar system against conventional electronic vital sign radars in terms of fractional bandwidth (the ratio between frequency tuning range and center frequency) and the system's range resolution. \Fref{FS7}\textbf{a} clearly shows that the demonstrated system has superior sensing bandwidth and range resolution compared with several existing vital sign radars; thus, better performance and accuracy for vital sign detection. It further proves that photonic radars have better frequency flexibility and tunability (fractional bandwidth) than conventional electronics. \Fref{FS7}\textbf{b} compares the photonics-based wideband RF signal generation in terms of the speed of the system's driving electronics and the demonstrated radar signal bandwidth across various photonic approaches, including using frequency-shifting cavity, laser sweeping, EOMs, and engineered fiber dispersion. It further proves that the demonstrated photonic radar based on the FS cavity requires simple electronics but sustains undiminished signal generation bandwidth, which reaches an optimal balance between the bandwidth limitation (the RF bandwidth of both the EOM and the driving electronics) from those based on EOMs and the bandwidth tuning flexibility of those based on dispersion.

\begin{table}[ht!]
    \centering
    \begin{tabularx}{0.95\textwidth}{ 
      >{\centering\arraybackslash}X 
      >{\centering\arraybackslash}X 
      >{\centering\arraybackslash}X
      >{\centering\arraybackslash}X
      >{\centering\arraybackslash}X
      >{\centering\arraybackslash}X
      >{\centering\arraybackslash}X
      >{\centering\arraybackslash}X} 
     \hline \hline
    Bandwidth (GHz) & Range resolution (mm) & Carrier Frequency (GHz) & Formality & Frequency tuning range (GHz) & Reference\\
    \end{tabularx}
    \begin{tabularx}{0.95\textwidth}{>{\centering\arraybackslash}X}
    \hline
    Photonic vital sign detection radar\\
    \end{tabularx}
    \begin{tabularx}{0.95\textwidth}{ 
      >{\centering\arraybackslash}X 
      >{\centering\arraybackslash}X 
      >{\centering\arraybackslash}X
      >{\centering\arraybackslash}X
      >{\centering\arraybackslash}X
      >{\centering\arraybackslash}X}
     \hline
     \textbf{10 (30)} & \textbf{13.74 (4.99)} & \textbf{30} & \textbf{SF} & \textbf{[MHz,>300]} &  \textbf{This Work}\\
     \hline
    \end{tabularx}
    \begin{tabularx}{0.95\textwidth}{>{\centering\arraybackslash}X}
    Electronic vital sign detection radar\\
    \end{tabularx}
    \begin{tabularx}{0.95\textwidth}{ 
      >{\centering\arraybackslash}X 
      >{\centering\arraybackslash}X 
      >{\centering\arraybackslash}X
      >{\centering\arraybackslash}X
      >{\centering\arraybackslash}X
      >{\centering\arraybackslash}X}
     \hline
     1 & 150 & 9 & SF & [8.5,9.5]  & \cite{Su2019}\\
     4 & 43 & 78-79 & LFM & [77,81]  & \cite{Ahmad2018}\\
     0.16 & 937.5 & 5.8 & LFM & [5.72,5.88]  & \cite{Wang2014}\\
     10 & 14.99$^{**}$ & 80 & LFM & [75,80]  & \cite{Wang2015}\\
     2 & 74.95$^{**}$ & 8 & SF & [7,9]  & \cite{Quaiyum2017}\\
     1.5 & 100 & 3 & FM & [2.25,3.75]  & \cite{Wang2013a}\\
     0.3 & 499.65$^{**}$ & 2.55 & SF & [2.4,2.7]  & \cite{Nahar2018}\\
     0.75 & 200 & 7.3 & LFM & [7.3,8.05]  & \cite{Mercuri2019}\\
     0.32 & 468.43$^{**}$ & 5.8 & CW+LFM & [5.64,5.96]  & \cite{Peng2017}\\
     - & - & 3 & CW & NA & \cite{Wang2013a}\\
     - & - & 5.8 & CW & NA & \cite{Tu2016}\\
     - & - & 2.4 & CW & NA & \cite{Wang2011}\\
     - & - & 60 & CW & [56,62] & \cite{Kao2012}\\
     - & - & 103 & CW & [93,105] & \cite{Ma2020}\\
    \hline
    \end{tabularx}
    \caption{\textbf{Comparison with reported electronic vital sign detection radars}. SF, stepped-frequency; LFM, linear-frequency modulated waveform; CW, continuous wave; NA, data not available or not mentioned in the literature; $^{**}$, range resolution not mentioned but calculated theoretically based on the reported bandwidth; -, not applicable.}
    \label{tab1}
\end{table}
\begin{table}[ht!]
    \centering
    \begin{tabularx}{0.95\textwidth}{ 
      >{\centering\arraybackslash}X 
      >{\centering\arraybackslash}X
      >{\centering\arraybackslash}X
      >{\centering\arraybackslash}X
      >{\centering\arraybackslash}X} 
     \hline \hline
    Demonstrated bandwidth (GHz) & Formality & linearity (\%)  & Mini. RF bandwidth (GHz)  & Reference\\
     \hline
     \textbf{30} & \textbf{SF} & \textbf{0.0036} & \textbf{0.1}  & \textbf{This Work}\\
     \hline
    \end{tabularx}
    \begin{tabularx}{0.95\textwidth}{>{\centering\arraybackslash}X}
    Photonic RF signal generation based on FS cavity\\
    \end{tabularx}
    \begin{tabularx}{0.95\textwidth}{ 
      >{\centering\arraybackslash}X 
      >{\centering\arraybackslash}X
      >{\centering\arraybackslash}X
      >{\centering\arraybackslash}X
      >{\centering\arraybackslash}X}
     \hline
     5.6 & SF & NA & 0.08  & \cite{Zhang2020a}\\
     26 & LFM,SFCS  & NA & 8  &\cite{Zhang2021}\\
     20 & SF  & NA & 0.08  & \cite{Lyu2022}\\
     27 & SF  & NA & 4  & \cite{Zhang2020c}\\
     >28 & LFM  & NA & 0.08  & \cite{DeChatellus2018}\\
    \hline
    \end{tabularx}
    \begin{tabularx}{0.95\textwidth}{>{\centering\arraybackslash}X}
    Photonic RF signal generation based on laser sweeping\\
    \end{tabularx}
    \begin{tabularx}{0.95\textwidth}{ 
      >{\centering\arraybackslash}X 
      >{\centering\arraybackslash}X
      >{\centering\arraybackslash}X
      >{\centering\arraybackslash}X
      >{\centering\arraybackslash}X}
     \hline
     4 & LFM  & NA & NA  & \cite{Zhou2022a}\\
     2 & LFM  & 7.8$^{**}$ & NA  & \cite{Zhang2020}\\
     7.5 & LFM  & NA & NA  & \cite{Hao2018}\\
     18.5 & LFM  & NA & 0.12  & \cite{Sun2021}\\
    \hline
    \end{tabularx}
    \begin{tabularx}{0.95\textwidth}{>{\centering\arraybackslash}X}
    Photonic RF signal generation based on dispersion\\
    \end{tabularx}
    \begin{tabularx}{0.95\textwidth}{ 
      >{\centering\arraybackslash}X 
      >{\centering\arraybackslash}X
      >{\centering\arraybackslash}X
      >{\centering\arraybackslash}X
      >{\centering\arraybackslash}X}
     \hline
     >30 & LFM & NA & NA &\cite{Li2014a}\\
    \hline
    \end{tabularx}
    \begin{tabularx}{0.95\textwidth}{>{\centering\arraybackslash}X}
    Photonic RF signal generation-based EOM\\
    \end{tabularx}
    \begin{tabularx}{0.95\textwidth}{ 
      >{\centering\arraybackslash}X 
      >{\centering\arraybackslash}X
      >{\centering\arraybackslash}X
      >{\centering\arraybackslash}X
      >{\centering\arraybackslash}X}
     \hline
     8 & LFM  & NA & 12 Gbit/s &\cite{Peng2018}\\
     6 & LFM  & NA & 9  &\cite{Li2020}\\
     8 & LFM  & NA & 6.5  & \cite{Zhang2017}\\
    \hline \hline
    \end{tabularx}
    \begin{tabularx}{0.95\textwidth}{>{\centering\arraybackslash}X}
    State-of-the-art AWG\\
    \end{tabularx}
    \begin{tabularx}{0.95\textwidth}{ 
      >{\centering\arraybackslash}X 
      >{\centering\arraybackslash}X
      >{\centering\arraybackslash}X
      >{\centering\arraybackslash}X
      >{\centering\arraybackslash}X}
     \hline
    25 & SF & 0.0013 & 25 & Keysight\\
    \hline
    \end{tabularx}
    \caption{\textbf{Comparison with reported photonic approaches.} SF, stepped-frequency; LFM, linear-frequency modulated waveform; CW, continuous wave; NA, data not available or not mentioned in the literature; $^{**}$, the linearity in the reference is the ratio between the maximum frequency deviation to the total bandwidth; -, not applicable.}
    \label{tab2}
\end{table}

\clearpage
\bibliography{supplement.bib}